\documentclass[secnumarabic,12pt,tightenlines,eqsecnum,floats,aps,amsmath,amssymb,nofootinbib,superscriptaddress,showpacs,prd]{revtex4-1}

\usepackage{amsmath,amsthm,amssymb,amsfonts}
\usepackage{graphicx}
\usepackage{palatino}
\usepackage{supertabular}
\usepackage[mathcal]{euler}
\usepackage{mathcomp}
\usepackage{float}
\usepackage{units}

\newenvironment{definition}[1][Definition]{\begin{trivlist}
\item[\hskip \labelsep {\bfseries #1}]}{\end{trivlist}}
 \newtheoremstyle{example}{\topsep}{\topsep}%
     {}
     {}
     {\bfseries}
     {}
     {\newline}
     {\thmname{#1}\thmnumber{ #2}\thmnote{ #3}}

   \theoremstyle{example}
   \newtheorem{example}{Example}[subsection]

\numberwithin{equation}{section}

\begin{document}

\title{The thermodynamic limit and black hole entropy in the area ensemble}

\author{J. Fernando \surname{Barbero G.}}
\affiliation{Instituto de
Estructura de la Materia, CSIC, Serrano 123, 28006 Madrid, Spain}

\author{Eduardo J. \surname{S. Villase\~nor}}
\affiliation{Instituto Gregorio
Mill\'an, Grupo de Modelizaci\'on y Simulaci\'on Num\'erica,
Universidad Carlos III de Madrid, Avda. de la Universidad 30, 28911
Legan\'es, Spain} \affiliation{Instituto de Estructura de la
Materia, CSIC, Serrano 123, 28006 Madrid, Spain}

\date{June 17, 2011}

\begin{abstract}
We discuss the thermodynamic limit in the canonical area ensemble used in loop quantum gravity to model quantum black holes. The computation of the thermodynamic limit is the rigorous way to obtain a smooth entropy from the counting entropy given by a direct determination of the number of microstates compatible with macroscopic quantities (the energy in standard statistical mechanics or the area in the framework presented here). As we will show in specific examples the leading behavior of the smoothed entropy for large horizon areas is the same as the counting entropy but the subleading contributions differ. This is important because these corrections determine the concavity or convexity of the entropy as a function of the area.
\end{abstract}

\pacs{04.70.Dy, 04.60.Pp}

\maketitle

\section{Introduction}

The standard formalism of thermodynamics can be based on the study of the entropy as the fundamental function encoding the full thermal behavior of a system. This point of view was actually advocated by some of the founding fathers of the subject, in particular by Gibbs \cite{Gibbs}. The entropy is a function of the equilibrium states of the system. For example, in the case of a gas, the equilibrium states are described by the total energy $E$ and the volume $V$ of the container. Hence, the entropy is a function of $(E,V)$. Other very important thermodynamical quantities are defined by derivatives of the entropy. In particular the inverse temperature and the pressure are given, respectively, by
$$
\frac{1}{T}=\frac{\partial S}{\partial E}\,,\quad \frac{P}{T}=\frac{\partial S}{\partial V}\,.
$$
The previous expressions only make sense if $S(E,V)$ satisfies some obvious and necessary smoothness conditions.

The entropy plays a central role in statistical mechanics. In fact, its determination by the counting of states subject to certain constraints (a fixed value of the energy, for example) can be taken as the starting point to understand the thermodynamical behavior of a system from the dynamics of its constituents. The actual result of such a counting usually suffers from a significant drawback: by its very nature it is always a staircase function of the natural variables describing the system (with jumps for the values of the basic variables for which the number of allowed states changes). This means that the derivatives necessary to compute relevant quantities, such as the temperature, are either zero or fail to be defined. A well-known solution to this problem goes back to the classical paper by Griffiths \cite{Griffiths} where the author shows how a smooth entropy  --with the possible exception of isolated points that would mark the existence of phase transitions--  can be found in a suitable \textit{thermodynamic limit} (see \cite{Fisher} and also \cite{Ruelle1999} for more details). This limit is defined by taking the size of the system to infinity while keeping some intensive parameters fixed. In the case of a gas, for example, the volume of its container is taken to infinity while keeping the number particle density and energy per particle fixed. In standard statistical mechanics texts this smoothing is sometimes carried out by considering discrete variables as continuous (which amounts in practice to some sort of coarse graining). Though this is not a rigorously defined procedure in many cases it provides the correct answer.

The identification of the laws of black hole mechanics \cite{Bardeen:1973gs} in classical general relativity suggested an unexpected connection between black hole physics and thermodynamics, in particular, a relationship between entropy and horizon area. The possibility of relating the horizon area with entropy was proposed by Bekenstein \cite{Bekenstein:1973ur} after the discovery by Hawking of the fact that the area of black hole never decreases \cite{Hawking:1971tu}. The exact proportionality between area and entropy, $S=A/4$ (in units $\hbar=c=G=k_B=1$), was fixed after the discovery of the Hawking radiation \cite{Hawking:1974sw} with a thermal spectrum.

The study of the microscopic origin of black hole entropy, i.e. black hole statistical mechanics, has received a lot of attention in the last fifteen years. The successful derivation of the Bekenstein-Hawking law $S=A/4$ has been claimed in several different approaches to quantum gravity and used by their proponents to highlight their relative merits. Two reasonably developed examples (but certainly not the only ones, see \cite{Wald:LR}) are provided by string theory and loop quantum gravity (LQG). In both cases there are schemes to define the entropy for certain types of black holes through the counting of microscopic configurations. Both approaches boast successful derivations of the Bekenstein-Hawking law.

The counting of states in string theory can be carried out for extremal \cite{Strominger:1996sh} or quasi extremal black holes (see \cite{Mandal:2010cj} for a recent review). Furthermore, the correspondence principle put forward by Horowitz and Polchinski \cite{Horowitz:1996nw} leads to a generic linear dependence between area and entropy for \textit{arbitrary} black holes though the proportionality constant cannot be obtained by invoking only this principle. These computations rely on the asymptotic determination of the number of D-brane configurations in the limit of large charges obtained by using the Cardy formula. The resulting expression for the number of states can be compared with the area of a classical black hole carrying the same charges. In the relevant limit (i.e. for large charges) one indeed finds the sought for linear relationship between entropy an area with the $1/4$ proportionality factor.

The loop quantum gravity framework approaches the same problem by modeling black holes with the help of the so called \textit{isolated horizons} \cite{Ashtekar:2004cn}. These can be understood as inner spacetime boundaries --with the topology $\mathbb{S}^2\times\mathbb{R}$-- that satisfy some geometrical conditions related to the presence of marginally trapped surfaces and energy conditions. An important result in this setting is the possibility of deriving laws of isolated horizon mechanics that mimic some of the laws of thermodynamics \cite{Ashtekar:1998sp}. The sector of general relativity consisting on space times with such inner boundaries admits a Hamiltonian formulation that can be taken as the starting point for its quantization \cite{Ashtekar:1999wa,Ashtekar:2000hw}. After introducing a suitable Hilbert space inspired in the ones customarily used in LQG (spanned by the so called spin network states) and solving the quantum constraint enforcing the isolated horizon conditions, it is possible to introduce a maximally degenerate density matrix $\rho$ and compute the quantum entropy as $\mathrm{Tr}\big(\rho\log\rho\big)$. When this is done a linear relationship between area and entropy is found for large horizon areas. The proportionality coefficient is a function of the so called Immirzi parameter $\gamma$ that must be then suitably fixed to recover the $1/4$ coefficient of the Bekenstein-Hawking law \cite{Ashtekar:2000eq,Engle:2009vc}. Although this is somewhat unsatisfactory, the choice of $\gamma$ is universal (in the sense that it is valid for all the types of black holed that have been studied) and hence the result can be used as a \textit{physical way} to fix this otherwise undetermined parameter. It is important to point out here that the LQG formalism can be applied to physical (i.e. non-extremal) black holes.

In the two settings considered above the \textit{counting entropy} (referred to by some authors as \textit{statistical entropy}) is a discontinuous function consisting in discrete steps. In order to get a suitable smooth function one has to consider the thermodynamic limit. It can be shown (by using, for example, a saddle point evaluation of certain integrals \cite{LopesCardoso:2006bg}) that the leading behavior of the true (smoothed) thermodynamical entropy coincides with the one corresponding to the statistical entropy. However, subdominant contributions can be different. This is specially important for black holes because the concavity or convexity of the entropy (related to the stability or lack thereof of the system) crucially depends on the behavior of these subdominant contributions.

The purpose of this paper is to discuss the issue of the thermodynamic limit for black holes. As we will show this is not simply the large area limit of the counting entropy. A direct and concrete derivation of this limit in the string framework has not been performed despite the fact that a very detailed knowledge of the microstates for some types of black holes has been recently obtained (see \cite{Mandal:2010cj} and references therein). The relevant counting of microscopic states in the LQG framework is also understood in great detail \cite{Agullo:2008yv,Agullo:2010zz,FernandoBarbero:2011kb}. In particular the microcanonical area  ensemble (the so called black hole degeneracy spectrum) is known in complete detail. Its Laplace transform defines the partition function in the canonical area ensemble; it can also be exactly obtained in this framework and used to illuminate some important features of the thermodynamic limit regarding, specifically, subdominant corrections to the asymptotic value of the entropy as a function of the area. In fact, this is the main goal of the paper. We will concentrate on the Domagala-Lewandowski (DL) \cite{Domagala:2004jt} and the Engle, Noui, Perez (ENP) \cite{Engle:2009vc} proposals (see also \cite{Engle:2010kt,Perez:2010pq}) but our results can be extended to other LQG inspired models. 

The main conclusion of our analysis is that the subdominant corrections to the smooth entropy obtained in the thermodynamic limit differ from the ones corresponding to the counting entropy. This result is actually expected because the theorems that guarantee the existence of the thermodynamic limit for the area ensemble show that the entropy must be concave. However it is in apparent conflict with the asymptotic behavior of the counting entropy (that has subdominant corrections to the area law that are proportional to \textit{minus} the logarithm of the area and are, hence, \textit{convex}).

The layout of the paper is the following. After this introduction we will briefly review in section \ref{Sect:2}  black hole and isolated horizon thermodynamics. We will then discuss in section \ref{Sect:3} the introduction of the canonical area ensemble and give the corresponding partition function. We will show that, in the thermodynamic limit, the entropy is a smooth and concave function of the area and we will determine asymptotic behavior for large areas. As we will see the linear behavior is the same as for the statistical entropy but the logarithmic corrections change. An important comment that is relevant at this point is the fact that the interesting structure \cite{Corichi:2006wn} found in the study of the black hole degeneracy spectrum cannot be present in the thermodynamical limit (though it may be relevant to study the detailed behavior of black holes in LQG and, in particular, Hawking radiation). We end the paper with our conclusions and comments and several appendices devoted to a brief review of the thermodynamic limit for some important sample systems and the asymptotic behavior of the partition functions.

\section{Black hole thermodynamics}\label{Sect:2}

The classic no-hair theorems tell us that black holes are described by a very small set of physical parameters: the mass $M$, the angular momentum $J$ and the electric charge $Q$ (see, however, \cite{Heusler:1998ua}). In particular the mass plays the role of the energy of a standard thermodynamical system. For Kerr-Newman black holes we have that the horizon area $A$, the surface gravity $\kappa$, the angular frequency of rotation $\Omega$ and the electrostatic potential $\Phi$ on the event horizon of the black hole are \cite{Hawking:1976de}
\begin{eqnarray*}
A(M,J,Q)&:=&\frac{4\pi}{M^2}\Big(J^2+(M^2+\sqrt{M^4-Q^2M^2-J^2})^2\Big)\,,\\
\kappa(M,J,Q)&:=&\frac{M\big(\sqrt{M^4-Q^2M^2-J^2}\big)}{J^2+(M^2+\sqrt{M^4-Q^2M^2-J^2})^2}\,,\\
\Omega(M,J,Q)&:=&\frac{4\pi J}{M\cdot A(M,J,Q)}\,,\\
\Phi(M,J,Q)&:=&\frac{4\pi Q (M^2+\sqrt{M^4-Q^2M^2-J^2})}{M\cdot A(M,J,Q)}\,.
\end{eqnarray*}
Those quantities satisfy the first law of black hole mechanics
\begin{equation}
\mathrm{d}M=\frac{\kappa}{8\pi}dA+\Omega \mathrm{d}J+\Phi \mathrm{d}Q\,.\label{firstlaw}
\end{equation}
If we define the black hole entropy and the temperature by
\begin{eqnarray*}
S(M,J,Q)&:=&\frac{A(M,Q,J)}{4}=\frac{\pi}{M^2}\Big(J^2+(M^2+\sqrt{M^4-Q^2M^2-J^2})^2\Big)\,,
\\
T(M,J,Q)&:=&\frac{\kappa(M,Q,J)}{2\pi}=\frac{1}{2\pi}\frac{M\big(\sqrt{M^4-Q^2M^2-J^2}\big)}{J^2+(M^2+\sqrt{M^4-Q^2M^2-J^2})^2}\,,
\end{eqnarray*}
the first law (\ref{firstlaw}) takes the usual form of the first law of thermodynamics
$$
\mathrm{d}M=T\mathrm{d}S+\Omega \mathrm{d}J+\Phi \mathrm{d}Q\,,
$$
where
$$
\frac{1}{T}=\frac{\partial S}{\partial M}\,,\quad \frac{\Omega}{T}=-\frac{\partial S}{\partial J}\,,\quad \frac{\Phi}{T}=-\frac{\partial S}{\partial Q}\,.
$$
Notice that the terms $\Omega \mathrm{d}J$ and $\Phi \mathrm{d} Q$ play the same role of $-P\mathrm{d}V$ for a gas, i.e. they are \textit{work} terms. It is important to realize that the area does not play the role of the volume.

Alternatively, it is also possible to express the mass (energy) as a function of the entropy and describe the black hole states in terms of $(S,J,Q)$. By using these variables and $A(S,J,Q)=4S$ we have now
\begin{eqnarray*}
M(S,J,Q)&=&\frac{\big((Q^2+S/\pi)^2+4J^2\big)^{\nicefrac{1}{2}}}{2(S/\pi\big)^{\nicefrac{1}{2}}}\,,\\
\kappa(S,J,Q)&=&2\pi\, T(S,J,Q)= \frac{(S/\pi)^2-(Q^4+4J^2)}{2(S/\pi)^{\nicefrac{3}{2}} \big((Q^2+S/\pi)^2+4J^2\big)^{\nicefrac{1}{2}}}\,,\\
\Omega(S,J,Q)&=&\frac{2J}{(S/\pi)^{\nicefrac{1}{2}}
\big((Q^2+S/\pi)^2+4J^2\big)^{\nicefrac{1}{2}}}\,,\\
\Phi(S,J,Q)&=&
\frac{(Q^2+S/\pi)Q}{(S/\pi)^{\nicefrac{1}{2}}\big((Q^2+S/\pi)^2+4J^2\big)^{\nicefrac{1}{2}}}\,.
\end{eqnarray*}
The temperature is defined, as usual, as the derivative of the energy with respect to the entropy at constant ``volume'' (i.e. constant $J$ and $Q$)
$$
T=\frac{\kappa}{2\pi}=\frac{\partial M}{\partial S}\,,\quad \Omega=\frac{\partial M}{\partial J}\,,\quad \Phi=\frac{\partial M}{\partial Q}\,.
$$

Isolated horizons  provide a generalization of the black hole event horizons. On one hand, the definition of isolated horizons is quasi-local, and it refers only to certain fields defined intrinsically on the horizon. On the other, only the intrinsic geometry of the isolated horizon is assumed to be time independent, whereas the geometry outside (not fixed by the one on the isolated horizon \cite{Lewandowski:1999zs}) may be non-stationary. In practice this means that there are nontrivial examples of isolated horizons in addition to the event horizons of the globally stationary black holes (which are isolated horizons themselves). As in the case of black holes, every state of an isolated horizon allows us to define the physical quantities $(M,J,Q)$ or, equivalently, $(S,J,Q)$. These numbers refer only to structures intrinsically defined on the horizon, without any reference to the behavior at infinity.\footnote{It is important to notice that, in contrast to black holes, the isolated horizon mass and
angular momentum do not suffice to provide a characterization the time independent horizon geometry. This characterization is given in terms of an infinite set of multipoles \cite{Ashtekar:2004gp} that capture the allowed distortions in the mass and angular momentum distribution on the horizon.} The first law of black hole mechanics can be generalized to the first law of isolated horizons and all formulas written above for Kerr-Newman black holes are still valid for (rigidly rotating) isolated horizons in the Einstein-Maxwell theory  \cite{Ashtekar:2001is,Ashtekar:2004cn}.

It is widely accepted that a successful quantum theory of gravity must explain the previous results from a microscopic point of view, however, there are some difficulties (beyond the obvious one of the lack of a fully working quantum gravity theory). For example, if the a microscopic description of a large mass Schwarzschild black hole is capable of explaining the relation $S(M,0,0)=4\pi M^2$, then the statistical mechanical density of energy microstates states has to satisfy $\log \Omega(M)\sim 4\pi M^2$ for large $M$. This means that the canonical ensemble is ill defined because \cite{Hawking:1976de}
$$
Z(\beta)=\int_0^\infty \Omega(M) e^{-\beta M} \mathrm{d}M \sim \int_0^\infty e^{4\pi M^2} e^{-\beta M}\, \mathrm{d} M
$$
diverges for all values of $\beta$. There are several possible ways to sidestep these problems. One can, for example, restrict oneself to working with the microcanonical (energy) ensemble. Another possibility is to put the back hole system inside a spherical cavity \cite{York:1986it}. Finally one can follow the suggestions of \cite{Krasnov:1996tb,Krasnov:1996wc} and use an \textit{area} canonical ensemble. This is the path that we will take here.

\section{Black hole area ensembles}\label{Sect:3}

The purpose of this section is to define and discuss the microcanonical and canonical area ensembles in LQG. The idea of using these types of ensembles goes back to Krasnov \cite{Krasnov:1996tb,Krasnov:1996wc,Krasnov:1997yt} and is, somehow, a necessity in the LQG formalism as the counting of states is naturally done by using the horizon area instead of black hole mass (see \cite{Ashtekar:1997yu,Ashtekar:2000eq}). In the following we will restrict ourselves to the spherically symmetric case but our results can be extended to more general situations such as the ones discussed in \cite{Ashtekar:2004nd}. A precursor of our work can be found in Meissner \cite{Meissner:2004ju}, where he solves the relevant counting problems by using Laplace transforms and, hence, he essentially derives the canonical ensemble (corresponding to the microcanonical area ensemble introduced by DL \cite{Domagala:2004jt} for the computation of the statistical entropy according to the recipe given in \cite{Ashtekar:2000eq}). Our presentation here ties a number of loose ends:

\begin{itemize}
\item We show how the exact resolution of the combinatorial problems \cite{Agullo:2008yv,G.:2008mj,Agullo:2010zz}, that gives the microcanonical ensemble and the statistical entropy, is a direct consequence of the use of Laplace transforms.
\item We argue that the canonical ensemble gives a smoothed and well behaved entropy. This is necessary to have the possibility of using the standard formalism of thermodynamics. An important consequence of deriving the entropy from the canonical ensemble is the fact that the subdominant corrections to the entropy \textit{do not} coincide with those corresponding to the statistical entropy. This result may be relevant outside the realm of LQG and should equally apply to string theory inspired models.
\item The smoothed entropy is \textit{concave} (i.e. the second derivative is negative) as a function of the area. This is actually a very general result (see \cite{Griffiths}) and is relevant to discuss the stability of the system. A word of caution may be necessary here because standard black holes are unstable. The likely reason behind this discrepancy is the use of the area ensemble. Actually, we are not claiming that the black holes are stable, but rather show that the use of the area ensemble has important physical consequences.
\item The entropy vanishes for zero area. This suggest a version of the third principle in the case of black holes in LQG.
\item The entropy given by the area canonical ensemble for a single black hole corresponds to the thermodynamical limit for a ensemble of non-interacting black holes (similar in spirit to the Einstein crystal model discussed in Appendix \ref{Sect:SimpleSystems}).

\end{itemize}

The main approaches to the problem of counting the configurations giving the statistical entropy in the microcanonical area ensemble for a back hole are those of DL \cite{Domagala:2004jt} and ENP \cite{Engle:2009vc}. For completeness we give the definitions of the statistical entropy in both cases.

\begin{definition}[Microscopic black hole entropy: DL-counting.]
According to Quantum Geometry and the Ashtekar-Baez-Corichi-Krasnov  framework, the counting entropy $S^{\scriptscriptstyle{\rm DL}}_{\mathrm{micro}}(A)$ of a quantum horizon of classical area $A$, is given by
$$S^{\scriptscriptstyle{\rm DL}}_{\mathrm{micro}}(A) = \log \Omega^{\scriptscriptstyle{\rm DL}}(A)\,,$$
where $\Omega^{\scriptscriptstyle{\rm DL}}(A)$ is one plus  the number of all the finite, arbitrarily long, sequences $(m_1,\ldots,m_N)$ of non-zero half integers, such that the following equality and inequality are satisfied:
$$\sum_{I=1}^N m_I=0, \quad \sum_{I=1}^N\sqrt{|m_I|(|m_I|+1)}\leq \frac{A}{8\pi\gamma\ell_P^2}.$$
The extra one in the definition of  $\Omega^{\scriptscriptstyle{\rm DL}}(A)$ comes from the trivial sequence.
\end{definition}

\noindent It is sometimes helpful to ignore the condition
$$
\sum_{I=1}^N m_I=0
$$
(the so called \textit{projection constraint}) to get a simplified entropy $S^{\scriptscriptstyle{\rm DL}}_{*\mathrm{micro}}(A) = \log \Omega^{\scriptscriptstyle{\rm DL}}_*(A)$ that is useful to understand some features of the entropy in the LQG framework. This has the effect of changing the subdominant terms in the asymptotic behavior of the entropy and, hence, contains important physics. This prescription must be understood as a way to rephrase the original counting problem in \cite{Ashtekar:2000eq} as one that can be solved by simpler methods.

\begin{definition}[Microscopic black hole entropy: ENP-counting (when $\gamma\leq\sqrt{3}$).]
The entropy $S^{\scriptscriptstyle{\rm ENP}}_{\rm micro}(A)$ of a quantum horizon of the classical area $A$  is defined as
$$S^{\scriptscriptstyle{\rm ENP}}_{\rm micro}(A) = \log  \Omega^{\scriptscriptstyle{\rm ENP}}(A)\,,$$
where $ \Omega^{\scriptscriptstyle{\rm ENP}}(A)$ is one plus the number of all the finite, arbitrarily long, sequences $(j_1,\ldots,j_N)$ of non-zero half integers $j_I$ satisfying
$$\sum_{I=1}^N\sqrt{j_I(j_I+1)}\leq \frac{A}{8\pi\gamma\ell^2_P}$$
and counted with a multiplicity given by the dimension of the invariant subspace $\mathrm{Inv}(\otimes_I[j_I])$.
\end{definition}

In principle the entropy defined above makes sense only for prequantized area values $A=A^{\rm CS}_k=4\pi\gamma\ell^2_P k$, $k\in\mathbb{N}$, however we extend the definition for arbitrary values of the area in the obvious way. In both cases the combinatorial problems involved in the computation of $\Omega(A)$ can be exactly solved by using number theoretic methods and the result conveniently encoded in generating functions as shown in \cite{Agullo:2008eg,BarberoG.:2008ue,Agullo:2009eq,Agullo:2010zz}. In the following we will use units of $4\pi\gamma\ell_P^2$

In the thermodynamic limit, and irrespective of the model, the  partition function per-particle $Z$ for an ensemble of non-interacting ``particles'' can be computed as the Laplace transform of the corresponding number of microstates $\Omega$ of a single object (see Appendix \ref{Sect:SimpleSystems}):
$$
\Omega(A)=\sum_{n=1}^\infty D_n \theta(A-A_n)\Leftrightarrow Z(\alpha)=\alpha\int_0^\infty e^{-\alpha A }\Omega(A)\,\mathrm{d}A=\sum_{n=1}^\infty D_n e^{-\alpha A_n }\,,
$$
where the integer numbers $D_n$ encode the black hole degeneracies associated with the area eigenvalues $A_n$ and $\theta$ denotes the Heaviside function. Notice that the parameter $\alpha$ is conjugate to the area and hence is not a temperature (which is conjugate to the energy).

In the thermodynamic limit (see Appendix \ref{Sect:SimpleSystems}), the average area is given by
$$
a(\alpha)=-\frac{\mathrm{d}}{\mathrm{d} \alpha}\log Z(\alpha)
$$
and the (smoothed) entropy can be computed as
$$
\tilde{\sigma}(\alpha):=\alpha a(\alpha)+\log Z(\alpha)\,.
$$
In practice, in order to express (and plot) the entropy as a function of the area, $a\mapsto\sigma(a)$, it is convenient to think of $\alpha$ as a parameter and consider the parametrized curve $\alpha\mapsto (a(\alpha),\tilde{\sigma}(\alpha))$.

In the DL scheme the partition functions $Z^{\scriptscriptstyle{\rm DL}}(\alpha)$ and $Z^{\scriptscriptstyle{\rm DL}}_*(\alpha)$ (with and without the projection constraint, respectively) can be read off directly from the integral expressions for the statistical entropy found in \cite{Meissner:2004ju,Agullo:2010zz}. When the projection constraint is not used $Z^{\scriptscriptstyle{\scriptscriptstyle{\rm DL}}}_*(\alpha)$ is given by
\begin{equation}
Z^{\scriptscriptstyle{\rm DL}}_*(\alpha)=\frac{1}{1-2\sum_{k=1}^\infty e^{-\alpha\sqrt{k(k+2)}}}\,.\label{ZDLstar}
\end{equation}
If the projection constraint is incorporated we have, instead
\begin{equation}
Z^{\scriptscriptstyle{\rm DL}}(\alpha)=\frac{1}{2\pi}\int_0^{2\pi} \frac{\mathrm{d}\omega}{1-2\sum_{k=1}^\infty e^{-\alpha\sqrt{k(k+2)}}\cos \omega k} \,.\label{ZDL}
\end{equation}
As real functions of the variable $\alpha$ both $Z^{\scriptscriptstyle{\rm DL}}(\alpha)$ and $Z^{\scriptscriptstyle{\rm DL}}_*(\alpha)$ have singularities for the unique value $\alpha^{\scriptscriptstyle{\rm DL}}_0\in\mathbb{R}$ satisfying
$$
1-2\sum_{k=1}^\infty e^{-\alpha^{\scriptscriptstyle{\rm DL}}_0\sqrt{k(k+2)}}=0\,.
$$
This can be directly seen in the case of $Z^{\scriptscriptstyle{\rm DL}}_*(\alpha)$, whereas for $Z^{\scriptscriptstyle{\rm DL}}(\alpha)$ the integral in the auxiliary variable $\omega$ diverges if $\alpha=\alpha^{\scriptscriptstyle{\rm DL}}_0=(0.746231\cdots)$ (and hence $Z^{\scriptscriptstyle{\rm DL}}(\alpha)$ is, itself, singular). This singularity controls the large area asymptotic behavior of the entropy whereas the asymptotic behavior of $Z(\alpha)$ in the regime  $\alpha\rightarrow \infty$ controls the limit $a\rightarrow 0$ of the entropy.

\begin{figure}[htbp]
\includegraphics[width=16cm]{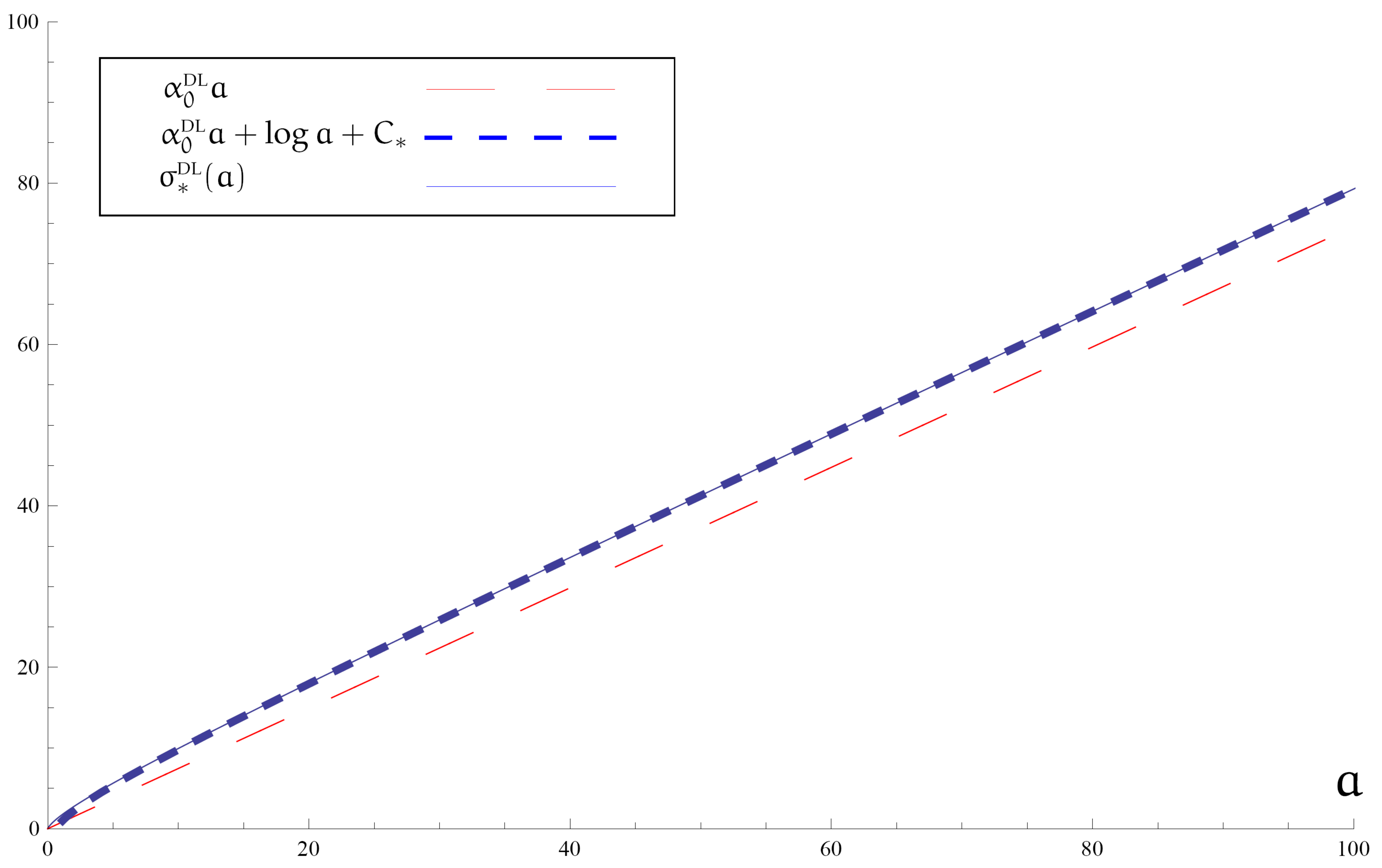}
\caption{Plot of the smoothed black hole entropy in the DL case without the projection constraint. We also plot the dominant asymptotic behavior that reproduces the Bekenstein-Hawking law and the correction obtained by considering the first subleading  terms $\log a+C_*$, with $C_*:=1+\log q^{\scriptscriptstyle{\rm DL}}_{*-1}$. Notice that both the exact entropy and its asymptotic approximation, $\alpha^{\scriptscriptstyle{\rm DL}}_0 a+\log a$, are convex. It is important to mention that no staircase structure appears and also that the entropy vanishes for zero area. The smoothed entropy is plotted by using the parametrization $\alpha\mapsto (a(\alpha),\tilde{\sigma}(\alpha))$ and numerically computing the partition function.} \label{Fig:1}
\end{figure}
\begin{figure}[htbp]
\includegraphics[width=16cm]{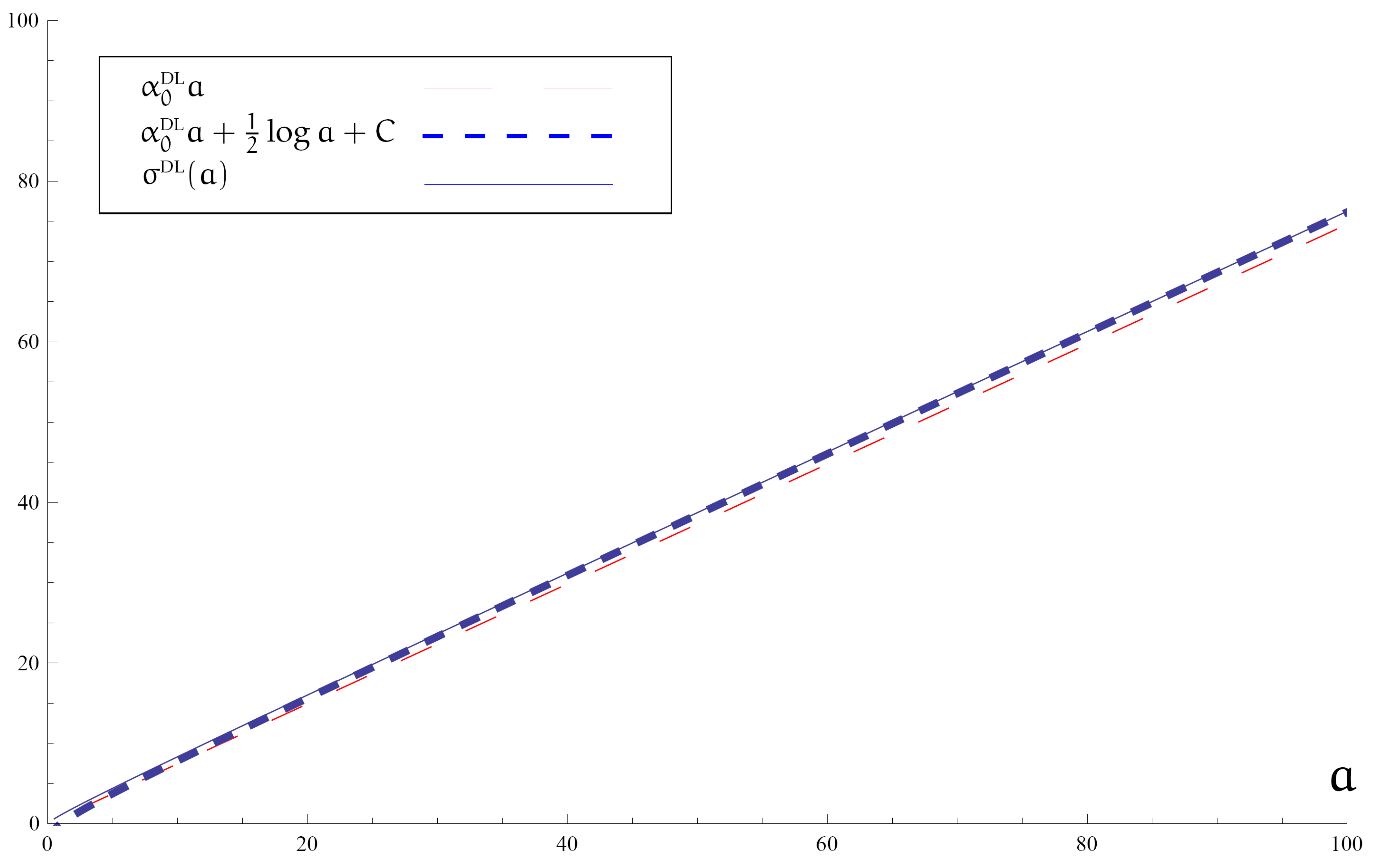}
\caption{Plot of the smoothed black hole entropy in the DL case when the projection constraint is taken into account (here $C:=\frac{1}{2}+\log q^{\scriptscriptstyle{\rm DL}}_{-\nicefrac{1}{2}}+\frac{1}{2}\log 2$). The main features are the same as in Fig. \ref{Fig:1}. The leading behavior is the same but the coefficient of the logarithmic term changes.} \label{Fig:2}
\end{figure}
\begin{figure}[htbp]
\includegraphics[width=16cm]{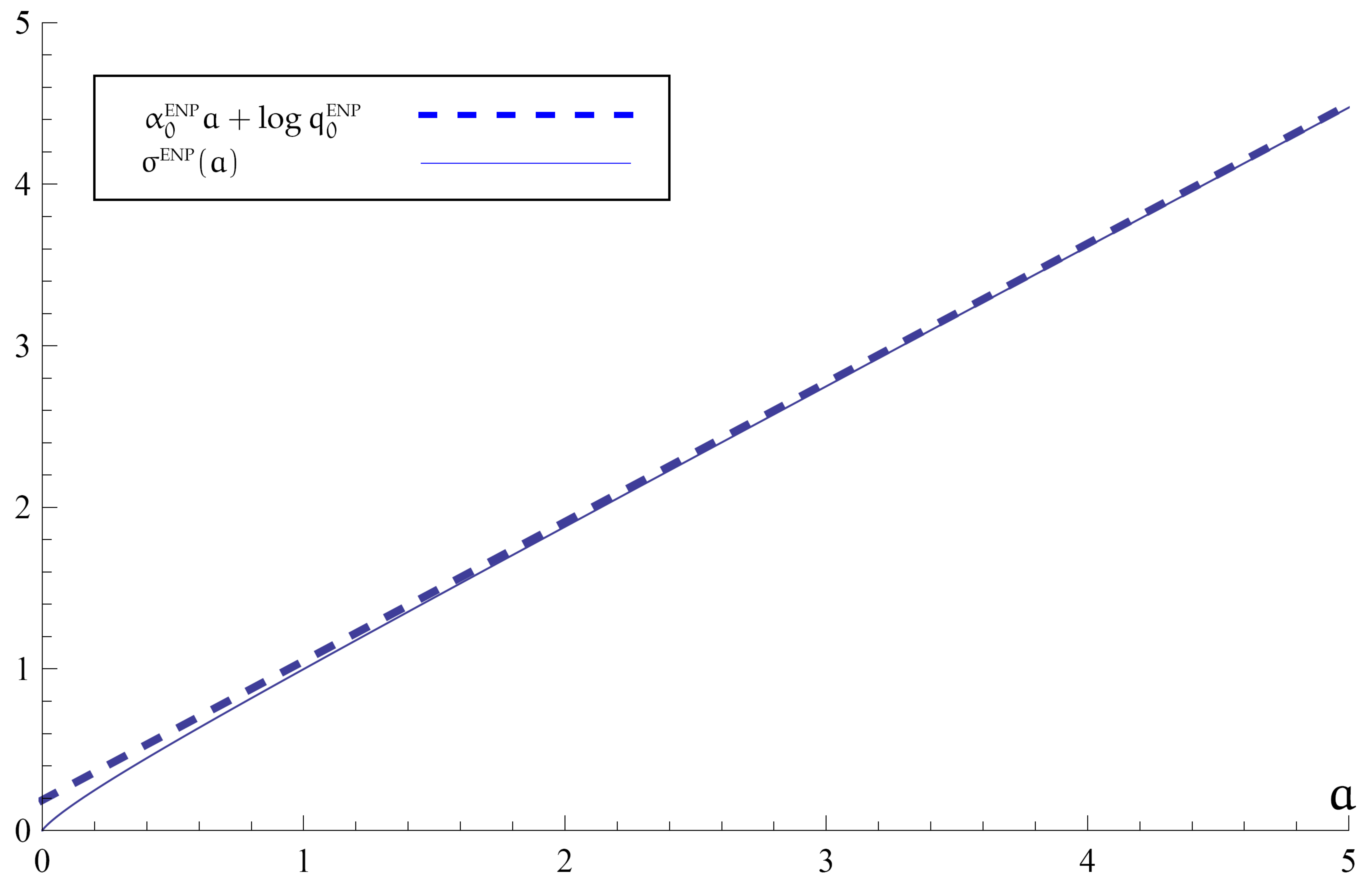}
\caption{Plot of the smoothed black hole entropy in the ENP case. We have plotted the entropy only to areas of around 5 (in units of $4\pi\gamma\ell_P^2$) because the linear regime is reached very early and the concavity would otherwise not be apparent.} \label{Fig:3}
\end{figure}

The asymptotic behaviors of $Z^{\scriptscriptstyle{\rm DL}}_*(\alpha)$ and $Z^{\scriptscriptstyle{\rm DL}}(\alpha)$ near $\alpha^{\scriptscriptstyle{\rm DL}}_0$ are,  respectively, of the form
\begin{eqnarray*}
Z^{\scriptscriptstyle{\rm DL}}_*(\alpha)&\sim& \frac{q^{\scriptscriptstyle{\rm DL}}_{*-1}}{\alpha-\alpha^{\scriptscriptstyle{\rm DL}}_0}+\sum_{n=0}^\infty q^{\scriptscriptstyle{\rm DL}}_{*n} (\alpha-\alpha^{\scriptscriptstyle{\rm DL}}_0)^n\,,\quad \alpha\rightarrow \alpha^{\scriptscriptstyle{\rm DL} +}_0\,,
\\
Z^{\scriptscriptstyle{\rm DL}}(\alpha)&\sim& \frac{q^{\scriptscriptstyle{\rm DL}}_{-\nicefrac{1}{2}}}{\sqrt{\alpha-\alpha^{\scriptscriptstyle{\rm DL}}_0}}+\sum_{n=0}^\infty q^{\scriptscriptstyle{\rm DL}}_{\nicefrac{n}{2}} (\alpha-\alpha^{\scriptscriptstyle{\rm DL}}_0)^{n/2}\,,\quad \alpha\rightarrow \alpha^{\scriptscriptstyle{\rm DL}+}_0\,,
\end{eqnarray*}
where the (non-zero) coefficients
\begin{eqnarray*}
q^{\scriptscriptstyle{\rm DL}}_{*-1}=\frac{1}{2\sum_{k=1}^\infty \sqrt{k(k+2)}e^{-\alpha^{\scriptscriptstyle{\rm DL}}_0\sqrt{k(k+2)}} }\,,\quad q^{\scriptscriptstyle{\rm DL}}_{-\nicefrac{1}{2}}=\frac{1}{2}\sqrt{\frac{q^{\scriptscriptstyle{\rm DL}}_{*-1}}{\sum_{k=1}^\infty k^2 e^{-\alpha^{\scriptscriptstyle{\rm DL}}_0\sqrt{k(k+2)}}}}\,.
\end{eqnarray*}
can be obtained in a straightforward way as discussed in Appendix \ref{App:Asymp}. The asymptotic behaviors of the entropy for large values of the area are then given by
\begin{eqnarray*}
\sigma^{\scriptscriptstyle{\rm DL}}_*(a)&\sim& \alpha^{\scriptscriptstyle{\rm DL}}_0 a+\log a+1+\log q^{\scriptscriptstyle{\rm DL}}_{*-1}+O(1/a)\,,\quad a\rightarrow\infty\,,\\
\sigma^{\scriptscriptstyle{\rm DL}}(a)&\sim& \alpha^{\scriptscriptstyle{\rm DL}}_0 a+\frac{1}{2}\log a+\frac{1}{2}+\log q^{\scriptscriptstyle{\rm DL}}_{-\nicefrac{1}{2}}+\frac{1}{2}\log 2+O(1/a)\quad a\rightarrow\infty\,.
\end{eqnarray*}
By fixing the Immirzi parameter $\gamma$ as in \cite{Meissner:2004ju} it is possible to recover the Bekenstein-Hawking area law.  Notice, however, that there are logarithmic corrections that imply that the difference between the entropy and the Bekenstein-Hawking law increases with the area, i.e. $\lim_{a\rightarrow \infty}(\sigma(a)-\alpha a)=\infty$. It is very important to notice that the corrections given above \textit{differ} from the ones corresponding to the statistical entropy \cite{Meissner:2004ju,Agullo:2010zz}
\begin{eqnarray*}
S^{\scriptscriptstyle{\rm DL}}_{*\mathrm{micro}}(A)\sim \alpha^{\scriptscriptstyle{\rm DL}}_0 A+o(A),\quad S^{\scriptscriptstyle{\rm DL}}_{\mathrm{micro}}(A)\sim \alpha^{\scriptscriptstyle{\rm DL}}_0 A-\frac{1}{2}\log A+o(\log A)\,.
\end{eqnarray*}
This difference is to be expected because the function $A\mapsto \alpha^{\scriptscriptstyle{\rm DL}}_0 A-\frac{1}{2}\log A$ is not concave.

Plots of the entropy as a function of the area and the asymptotic approximations can be seen in Figs. \ref{Fig:1} and \ref{Fig:2}. One can readily see that the entropy is a concave (negative second derivative) function of the area. It is also worthwhile to notice how fast the entropy approaches its asymptotic approximation (a concave function itself in the DL case).

The partition function in the ENP case can be obtained from the results of \cite{Agullo:2009eq}, it is given by
\begin{eqnarray*}
Z^{\scriptscriptstyle{\rm ENP}}(\alpha)=\frac{1}{\pi}\int_0^{2\pi}\frac{\sin^2 \omega\,\mathrm{d}\omega}{1-\sum_{k=1}^\infty e^{-\alpha \sqrt{k(k+2)}}\sin\big( (k+1)\omega\big)/\sin\omega}\,.
\end{eqnarray*}
The singularity $\alpha^{\scriptscriptstyle{\rm ENP}}_0=(0.861006\cdots)$ in the partition function that dictates the large area behavior of the entropy is defined now by
$$
1-\sum_{k=1}^\infty (k+1)e^{-\alpha^{\scriptscriptstyle{\rm ENP}}_0\sqrt{k(k+2)}}=0\,.
$$
In the vicinity of this singularity the partition function behaves as (see Appendix \ref{App:Asymp})
\begin{eqnarray*}
Z^{\scriptscriptstyle{\rm ENP}}(\alpha)\sim q_0^{\scriptscriptstyle{\rm ENP}}+q^{\scriptscriptstyle{\rm ENP}}_{\nicefrac{1}{2}}\sqrt{\alpha-\alpha^{\scriptscriptstyle{\rm ENP}}_0}+\sum_{n=2}^\infty q^{\scriptscriptstyle{\rm ENP}}_{\nicefrac{n}{2}}(\alpha-\alpha^{\scriptscriptstyle{\rm ENP}}_0)^{n/2}
\end{eqnarray*}
and the asymptotic behavior of the entropy is
\begin{eqnarray*}
\sigma^{\scriptscriptstyle{\rm ENP}}(a)&\sim& \alpha^{\scriptscriptstyle{\rm ENP}}_0 a+\log q_0^{\scriptscriptstyle{\rm ENP}}+ O(1/a)\,,\quad a\rightarrow \infty\,.
\end{eqnarray*}
As in the DL case, the vale of $\gamma$ can be fixed to recover  the Bekenstein-Hawking \cite{Agullo:2009eq}. Notice that for the ENP model there are no logarithmic corrections (only the constant term $\log q_0^{\scriptscriptstyle{\rm ENP}}$) and, hence, $\sigma^{\scriptscriptstyle{\rm ENP}}(a)- \alpha^{\scriptscriptstyle{\rm ENP}}_0 a$ remains finite when $a\rightarrow\infty$.

In this case, the correction also differs from the ones corresponding to the statistical entropy \cite{Agullo:2009eq}
\begin{eqnarray*}
S^{\scriptscriptstyle{\rm ENP}}_{\mathrm{micro}}(A)\sim \alpha^{\scriptscriptstyle{\rm ENP}}_0 A-\frac{3}{2}\log A+o(\log A)\,.
\end{eqnarray*}

\section{Comments}

The main point that we make in the paper is that it is not enough to determine the statistical entropy of a black hole by counting its microscopical states but, in addition, one should go to the thermodynamic limit in an appropriate way. This is unavoidable in order to find a smooth entropy function that can be differentiated to compute physical quantities according to the well established rules of thermodynamics. Smoothness is related to convexity properties that are relevant to understand stability in thermodynamical terms and define the single valued Legendre transforms used in the definition of thermodynamic potentials. It is only in this limit that the eventual singularities of the entropy can be used to identify and study phase transitions. These issues are relevant if one is going to be serious about black hole thermodynamics.

An illustration of the kind of problems that can be expected has been discussed here in the context of entropy computations for black holes in loop quantum gravity. The discussion is facilitated by the availability of very detailed counting schemes that enable a precise counting of states as functions of the horizon area. Actually it is possible to obtain the (area) partition function for the system  in closed form  by using generating functions (as hinted in the paper by Meissner \cite{Meissner:2004ju}). With the help of the partition function it is straightforward to derive the form of the entropy in the thermodynamic limit.

The methods that we have used through the paper are based on very general results in statistical mechanics. This means that we can extend our work to other LQG inspired approaches in addition to the ones discussed in this paper. In particular, it is possible to do it for the modified flux area operator model with equally spaced area eigenvalues \cite{FernandoBarbero:2009ai} with conclusions similar to the ones found here. The discussion presented in the paper should also be relevant for the proposals appearing in \cite{Sahlmann:2011xu} and \cite{Krasnov:2009pd}.

It is important to point out here that the thermodynamic limit is not just the limit of large areas. This is so for several reasons.
\begin{itemize}
\item The large area limit as such does not lead to a well defined smooth or concave entropy function. A concrete smoothing procedure must be implemented.
\item The role of the area as a thermodynamical variable for black holes must be carefully understood. In the fundamental example provided by the Schwarzschild black hole the correct way to interpret the results on the entropy is to state that it is a function of the mass (energy) $S(M)=4\pi M^2$. The horizon area is $A(M)=16\pi M^2$ and the temperature is $T(M)=1/(8\pi M)$. It is true that the entropy is proportional to the horizon area (i.e. entropy and area are not independent) but, in order to get the temperature by the using the standard rules of thermodynamics the entropy must be expressed as a function of the mass.
\item It is not clear at all that the black hole area plays the same role as the volume of a gas. This means that it is not obvious what kind of intensive parameter should be kept fixed when defining the thermodynamic limit (number of punctures divided by horizon area?). This problem disappears (or at least takes a different disguise) if one considers an ensemble of ``independent'' black holes similar to the Einstein model for a crystal. Although this leads to a definite prescription to get smooth thermodynamical properties one would expect the actual microscopic gravitational degrees of freedom and their interactions to play a relevant role (as spin interactions do in a ferromagnet). Our examples are meant to illustrate the importance of considering the thermodynamic limit but the ignorance about the exact dynamics of the system prevents us from going further.
\item Finally, as we have shown, the subleading corrections do not necessarily coincide for the smoothed entropy derived in the thermodynamic limit and for the statistical one. Care must be duly exercised then when comparing subleading contributions to the statistical entropy with their ``macroscopic'' counterparts (as is frequently done in the literature).
\end{itemize}

We want to add several more comments. First we want to emphasize the fact that the mere use of density matrices to describe quantum black holes means that we are actually working with some kind of statistical ensemble. In practice we are forced to measure a large number of times on a system consisting of a single object and prepared according to some concrete prescription.  Nobody would hesitate, in the context of quantum mechanics, to talk about the density matrix for say a hydrogen atom as a way to encode the classical uncertainty involved in the incomplete preparation of such a system. The way entropy is defined in LQG relies on a ``maximal degeneracy'' density matrix and, hence, incorporates this type of uncertainty. The Einstein crystal-like model that we are using to introduce a thermodynamic limit is inspired by this point of view.

There is a certain ambiguity in the definition of the number of microstates as a consequence of the alternative ways to think about the system. For example, in the case of LQG black holes one can wonder whether the different sets of punctures (with their labels) are the analogues of the energy levels of a harmonic oscillator or if they should be considered as the particles in a gas (in an analogous way, one can wonder if the different D-brane configurations for a black hole system in string theory are really its microscopic constituents or they should be thought of as the levels in a harmonic oscillator).

As a final comment we want to mention the fact that one should really work with an energy operator instead of an area operator. Although this  is not available in the LQG models that we are using, one could try to postulate it [employing, for example the standard one for Schwarzschild black holes $M=\sqrt{A}/(4\sqrt{\pi}) $] and deal with the lack of a well defined canonical ensemble by working directly with the microcanonical one in order to go to the thermodynamic limit. The results derived with the canonical area ensemble cannot be directly generalized to get the solution to this problem but on the other hand we expect that the general results that we have found (in particular the different subleading behaviors of the statistical entropy and the smooth entropy in the thermodynamic limit) will be generally true.

\begin{acknowledgements}
We want to thank Ivan Agullo, Enrique F. Borja, Alejandro Corichi, Jacobo Diaz-Polo, Alejandro Perez, Hanno Sahlmann and Madhavan Varadarajan for interesting discussions and comments. This work has been supported by the Spanish MICINN research grant FIS2009-11893 and  the  Consolider-Ingenio 2010 Program CPAN (CSD2007-00042).
\end{acknowledgements}

\appendix

\section{The thermodynamical limit for simple systems.}{\label{Sect:SimpleSystems}}

In this appendix we briefly review the thermodynamic limit for some important sample systems (see \cite{Fisher,Griffiths,Ruelle1999,Ashcroft} for more details).

The quantum description of a $N$-particle system confined in the domain $\Lambda\subset \mathbb{R}^3$ is provided by a Hamiltonian operator $\mathbf{H}(N;\Lambda)$. Let
$$\Omega(E,N;\Lambda)= \mathrm{Tr}\Big(\theta\big(E-\mathbf{H}(N;\Lambda)\big)\Big)$$
be the number of eigenstates of $\mathbf{H}(N;\Lambda)$ not exceeding $E$. The counting entropy in the quantum microcanonical ensemble is defined as
$$S_{\mathrm{micro}}(E,N;\Lambda)=\log \Omega(E,N;\Lambda)\,.$$
As a function of the energy (taken as a real variable) it is a staircase function. This means, in particular, that it is either non-differentiable or the derivative is zero, a fact that precludes the use of the standard formulas in thermodynamics.
In the thermodynamic limit, the entropy per-volume is given by
$$
\sigma(\epsilon,\rho)=\lim_{V(\Lambda)\rightarrow\infty} \frac{S_{\mathrm{micro}}( \epsilon V(\Lambda),\rho V(\Lambda); \Lambda)}{V(\Lambda)}\,,
$$
where $V(\Lambda)$ is the volume of the domain $\Lambda$, $\epsilon$ is the energy per volume and $\rho$ the number density. In some cases, the role of $V(\Lambda)$ can be played by other extensive parameters (for example by the number of particles itself).  With the exception of a set of zero measure (that corresponds to phase transitions) the function $\sigma$ satisfies some regularity properties, in particular it is concave as a function of the energy $\epsilon$ and, consequently, it is differentiable. This allows us to define state functions such as the (inverse) temperature
$$
\beta(\epsilon,\rho)=\frac{1}{T(\epsilon,\rho)}=\frac{\partial \sigma}{\partial \epsilon}(\epsilon,\rho)\,.
$$

In the quantum canonical ensemble the temperature characterizes a thermal bath with which the thermodynamical system interacts (exchanging energy). In this scheme the fundamental object that contains the relevant information about the system is the partition function
$$
Z_N(\beta,\Lambda)=\mathrm{Tr}\Big(\exp\big(-\beta \mathbf{H}(N;\Lambda)\big)\Big)
$$
or, equivalently, the free energy
$$
F_N(\beta,\Lambda)=\log Z_N(\beta,\Lambda)\,.
$$
In the thermodynamic limit, the free energy per volume is
$$
f(\beta,\rho)=\lim_{\Lambda\rightarrow \infty}\frac{F_{N(\Lambda)}(\beta,\Lambda)}{V(\Lambda)}=\lim_{\Lambda\rightarrow \infty} \frac{\log Z_{N(\Lambda)}(\beta,\Lambda)}{V(\Lambda)}\,, \textrm{ where } \rho=\frac{N(\Lambda)}{V(\Lambda)}\,.
$$
In this scheme the energy per volume is a derived quantity given by
$$
\epsilon(\beta,\rho)=-\frac{\partial f}{\partial \beta}(\beta,\rho)
$$
and the entropy per particle is obtained by a Legendre transform
$$
\tilde{\sigma}(\beta,\rho)=\beta\epsilon(\beta,\rho)+f(\beta,\rho)\,.
$$
The existence of the Legendre transform is guaranteed now by the convexity properties of the free energy defined in the thermodynamic limit. Finally it is important to point out that for reasonable interactions the thermodynamic limit is the same for both the microcanonical and canonical ensembles, in particular we have
$$
\sigma(\epsilon,\rho)=\tilde{\sigma}(\beta(\epsilon,\rho),\rho)\,.
$$
We give now some examples relevant for the discussion in the main body of the paper.

\begin{example}[One-dimensional Einstein crystal.] Let us consider a system consisting in $N$ non-interacting one dimensional oscillators. The Hamiltonian is then  $\mathbf{H}(N)=\oplus_{i=1}^N H_i$, where each $H_i=\omega a^*_ia_i$ is a (normal ordered) one-dimensional harmonic oscillator. For this system, it is straightforward tho show that in the quantum microcanonical ensemble
\begin{eqnarray*}
\Omega(E,N)&=&\sum_{n=0}^{\lfloor E/\omega \rfloor} \binom{N-1+n}{N-1}=\binom{N+\lfloor E/\omega  \rfloor}{N}\,,\\ S_{\mathrm{micro}}(E,N)&=&\log\Omega(E,N)=\log \binom{N+\lfloor E/\omega  \rfloor}{N}\,.
\end{eqnarray*}
It is important to notice that, irrespective of the number of oscillators $N\in \mathbb{N}$, the counting entropy $E\mapsto S_{\mathrm{micro}}(E,N)$ is a staircase function of $E\in [0,\infty)$. However, in the thermodynamic limit the thermodynamic entropy $\epsilon \mapsto \sigma(\epsilon)$, defined as
\begin{eqnarray}
\sigma(\epsilon)&=&\lim_{N\rightarrow \infty} \frac{S_{\mathrm{micro}}(N\epsilon,N)}{N}=\lim_{N\rightarrow \infty}\frac{1}{N}\log \binom{N+\lfloor N\epsilon/\omega  \rfloor}{N}\nonumber\\
&=&\frac{\epsilon}{\omega} \log(1+\omega/\epsilon)+\log(1+\epsilon/\omega)\,,\label{S}
\end{eqnarray}
is a smooth, concave, function for $\epsilon \in [0,\infty)$ that satisfies $\sigma(0)=0$\,. The temperature is a derived quantity
\begin{eqnarray*}
\beta(\epsilon)=\frac{d\sigma}{d\epsilon}(\epsilon)=\frac{\log(1+\omega/\epsilon)}{\omega}\,.
\end{eqnarray*}

On the other hand, if we take the thermodynamic limit within the canonical ensemble point view, the thermodynamic free energy per particle $f$ coincides with the logarithm of the partition function of a single quantum harmonic oscillator, i.e.
$$
f(\beta)=\lim_{N\rightarrow \infty}\frac{F_N(\beta)}{N}=F_1(\beta)=\log\Big(Tr\big(\exp(-\beta H_1)\big)\Big)=\log\left(\sum_{n=0}^\infty e^{-n\beta\omega}\right)=\log\left(\frac{1}{1-e^{-\beta\omega}}\right)\,.
$$
In this approach the mean energy per particle, $\epsilon(\beta)$, is
\begin{eqnarray*}
\epsilon(\beta)=-\frac{d f}{d\beta}(\beta)=\frac{\omega}{e^{\beta\omega}-1}\label{EdeT}\,.
\end{eqnarray*}
Finally the entropy is computed as a Legendre transform
\begin{equation}
\tilde{\sigma}(\beta)= \beta\epsilon(\beta)+ f(\beta)=\frac{\beta\omega}{e^{\beta\omega}-1}-\log(1-e^{-\beta\omega})\,.\label{TildeS}
\end{equation}
As expected, (\ref{S}) and (\ref{TildeS}) are equivalent
$$
\sigma(\epsilon)=\tilde{\sigma}(\beta(\epsilon))\,,\quad \tilde{\sigma}(\beta)=\sigma(\epsilon(\beta))\,.
$$
An interesting fact, relevant in this example and due to the non-interacting nature of its constituents, is the possibility of finding the explicit relationship between the true and the statistical entropy for a single object. If we do not take the thermodynamic limit in the microcanonical ensemble and consider the $N=1$ case, denoting $\Omega_1(E):=\Omega(E,N=1)$ and $S_1(E)=S_{\mathrm{micro}}(E,N=1)$, we get
$$
\Omega_1(E)=\exp S_1(E)=\binom{1+\lfloor E/\omega \rfloor}{1}=1+\lfloor E/\omega \rfloor =\sum_{n=0}^\infty \theta(E/\omega-n)
$$
The Laplace transform $\mathcal{L}$ of $\exp S_1$, as a function of $\beta$, is given by
$$
\mathcal{L}(\Omega_1, \beta)=\mathcal{L}(\exp S_1, \beta)=\frac{1}{\beta}\sum_{n=0}^\infty e^{-n\beta\omega}=\frac{1}{\beta}\frac{1}{1-e^{-\beta\omega}}\,.
$$
Hence, as expected, the canonical partition function of a single quantum harmonic oscillator is, essentially,  the Laplace transform of the exponential of the counting entropy in the microcanonical esemble
\begin{eqnarray}
Z_1(\beta)= \beta \mathcal{L}(\Omega_1, \beta)=\beta \mathcal{L}(\exp S_1, \beta)\,.\label{Laplace}
\end{eqnarray}
The free energy per particle in the thermodynamic limit in the canonical ensemble is given by
$$
f(\beta)=\log Z_1(\beta)=\log (\beta\mathcal{L}(\exp S_1, \beta))
$$
and
$$
\epsilon(\beta)=-\frac{\mathrm{d}}{\mathrm{d}\beta} \log (\beta \mathcal{L}(\exp S_1, \beta))
$$
Finally the relationship between both entropies is
$$
\tilde{\sigma}(\beta)=-\beta\frac{\mathrm{d}}{\mathrm{d}\beta}\log (\beta \mathcal{L}(\exp S_1, \beta))+\log (\beta\mathcal{L}(\exp S_1, \beta))
$$

\end{example}

\begin{example}[One-dimensional periodic lattice (phonons).]

As it is well known, (see, for example, \cite{Ashcroft}), the normal modes of a one-dimensional periodic lattice are described by the (normal ordered) Hamiltonian
$$
\mathbf{H}(N)=\sum_{n=1}^N \omega_n(N) a^*_n a_n
$$
where
$$
\omega_n(N)=2 \omega_0 \left|\sin\left(\frac{c k_n(N)}{2}\right)\right|\,,\quad k_n(N)=\frac{2\pi n}{Nc}\,,\quad n\in{1,\dots,N}\,,
$$
and $c$ is the lattice constant. The partition function is given by
$$
Z_N(\beta)=\prod_{n=1}^N\frac{1}{1-\exp(-\beta \omega_n(N))}
$$
and hence, in the thermodynamic limit, the free energy (per unit length) is
\begin{eqnarray*}
f(\beta)&=&\lim_{N\rightarrow \infty} \frac{1}{Nc}\sum_{n=1}^N\log\left(\frac{1}{1-\exp(-\beta\omega_n(N))}\right)\\&=&\frac{1}{2\pi}\int_0^{2\pi/c}
\log\left(\frac{1}{1-\exp(- 2 \beta  \omega_0 \sin (ck/2) )}\right)\,\mathrm{d}k\,.
\end{eqnarray*}
\end{example}

\begin{example}[Black body radiation (photons)]
In this case, the extensive parameter that we will use in the thermodynamic limit is $V=L^3$. The Hamiltonian is now
$$
\mathbf{H}(L)=\sum_{\mathbf{n}\in \mathbb{Z}^3} \sum_{\lambda=\pm } \omega_{\mathbf{n}}(L) a^*_{\mathbf{n},\lambda}(L) a_{\mathbf{n},\lambda}(L)\,,
$$
where $\lambda=\pm $ are the two polarizations of the photon,  $\mathbf{n}=(n_1,n_2,n_2)\in \mathbb{Z}^3$ and
$$
\omega_{\mathbf{n}}(L):=\sqrt{\mathbf{k}_{\mathbf{n}}^2(L)}\,\quad \mathbf{k}_{\mathbf{n}}(L)=\frac{2\pi}{L} \mathbf{n}\,.
$$
The partition function for finite volume is the product of the partition function corresponding to each normal mode
$$
Z_L(\beta)=\prod_{\mathbf{n}\in \mathbb{Z}^3}\left(\frac{1}{1-\exp\big(-\beta \omega_{\mathbf{n}}(L)\big)}\right)^2\,.
$$
In the thermodynamic limit the free energy per volume is
\begin{eqnarray*}
f(\beta)&=&\lim_{L\rightarrow \infty}\frac{\log Z_L(\beta)}{L^3}=\lim_{L\rightarrow \infty} \frac{2}{L^3}\sum_{\mathbf{n}\in \mathbb{Z}^3}\log\left(\frac{1}{1-\exp\big(-\beta \omega_{\mathbf{n}}(L)\big)}\right)
\\
&=& \frac{2}{(2\pi)^3}\int_{\mathbb{R}^3} \log\left(\frac{1}{1-e^{-\beta \sqrt{\mathbf{k}^2}}}\right)  \mathrm{d}^3\mathbf{k}
=
\frac{1}{\pi^2}\int_0^\infty  \log\left(\frac{1}{1-e^{-\beta \omega}}\right) \omega^2 \mathrm{d}\omega=\frac{\pi^2}{45\beta^3}\,,
\end{eqnarray*}
and the energy density (energy per unit of volume) of the radiation is given by
$$
\epsilon(\beta)= -\frac{df}{d\beta}(\beta)=
\frac{1}{\pi^2}\int_0^\infty \frac{\omega^3d\omega}{e^{\beta \omega}-1}=\frac{\pi^2}{15\beta^4}\,.
$$
The entropy as a function of the temperature is
$$
\tilde{\sigma}(\beta)=\beta \epsilon(\beta)+f(\beta)=\frac{4\pi^2}{45 \beta^3}\,,
$$
hence, the entropy is a concave function of the energy
$$
\sigma(\epsilon)=\frac{4}{3}\frac{\pi^{\nicefrac{1}{2}}}{15^{\nicefrac{1}{4}}}\,\epsilon^{3/4}\,.
$$
\end{example}

\section{Asymptotic behavior of the partition functions}{\label{App:Asymp}}
Let us show first that
$$
Z^{\scriptscriptstyle{\rm DL}}_*(\alpha)\sim \frac{q^{\scriptscriptstyle{\rm DL}}_{*-1}}{\alpha-\alpha^{\scriptscriptstyle{\rm DL}}_0}+\sum_{n=0}^\infty q^{\scriptscriptstyle{\rm DL}}_{*n} (\alpha-\alpha^{\scriptscriptstyle{\rm DL}}_0)^n\,,\quad \alpha\rightarrow \alpha^{\scriptscriptstyle{\rm DL} +}_0\,.
$$
As $Z^{\scriptscriptstyle{\rm DL}}_*(\alpha)$, defined in (\ref{ZDLstar}), is a meromorphic function the relevant coefficients $q_{*n}^{\scriptscriptstyle{\rm DL}}$ can be simply obtained by computing limits when $\alpha\rightarrow \alpha^{\scriptscriptstyle{\rm DL}}_0$. In particular, to the order that we are considering in the text of the paper, we only need to know
$$
q_{*-1}^{\scriptscriptstyle{\rm DL}}=\lim_{\alpha\rightarrow \alpha^{\scriptscriptstyle{\rm DL}}_0}(\alpha-\alpha^{\scriptscriptstyle{\rm DL}}_0)Z_*^{\scriptscriptstyle{\rm DL}}(\alpha)=\frac{1}{2\sum_{k=1}^\infty \sqrt{k(k+2)}e^{-\alpha^{\scriptscriptstyle{\rm DL}}_0\sqrt{k(k+2)}} }\,.
$$

The expression for $Z^{\scriptscriptstyle{\rm DL}}(\alpha)$ given in (\ref{ZDL}) involves an integral so, in this case, we cannot proceed as before and, in fact, we actually need to find the asymptotic behavior of the integral in the limit $\alpha\rightarrow \alpha^{\scriptscriptstyle{\rm DL} +}_0$. The best strategy in this case consists in writing
$$
Z^{\scriptscriptstyle{\rm DL}}(\alpha)=\frac{1}{2\pi}\int_{-\pi}^\pi\frac{\mathrm{d}\omega}{Q^{\scriptscriptstyle{\rm DL}}(\alpha,\omega)}
$$
with
$$
Q^{\scriptscriptstyle{\rm DL}}(\alpha,\omega):=1-2\sum_{k=1}^\infty e^{-\alpha\sqrt{k(k+2)}}\cos\omega k=\Big(1-2\sum_{k=1}^\infty e^{-\alpha\sqrt{k(k+2)}}\Big)+\Big(4\sum_{k=1}^\infty e^{-\alpha\sqrt{k(k+2)}}\sin^2\frac{\omega k}{2}\Big)\,.
$$
We define now
\begin{eqnarray*}
Q^{\scriptscriptstyle{\rm DL}}_0(\alpha,\omega)&:=&\Big(1-2\sum_{k=1}^\infty e^{-\alpha\sqrt{k(k+2)}}\Big)+\Big(4\sum_{k=1}^\infty k^2e^{-\alpha\sqrt{k(k+2)}}\Big)\sin^2\frac{\omega}{2}\\
&=:&C^2(\alpha)+B^2(\alpha)\sin^2\frac{\omega}{2}
\end{eqnarray*}
and
$$
G^{\scriptscriptstyle{\rm DL}}(\alpha,\omega):=Q^{\scriptscriptstyle{\rm DL}}_0(\alpha,\omega)-Q^{\scriptscriptstyle{\rm DL}}(\alpha,\omega)=4\sum_{k=2}^\infty e^{-\alpha\sqrt{k(k+2)}}\Big(k^2\sin^2\frac{\omega}{2}-\sin^2\frac{\omega k}{2}\Big)\,,
$$
to write
$$
\frac{1}{Q^{\scriptscriptstyle{\rm DL}}}=\frac{1}{Q^{\scriptscriptstyle{\rm DL}}_0}+\left( \frac{1}{Q^{\scriptscriptstyle{\rm DL}}}-\frac{1}{Q^{\scriptscriptstyle{\rm DL}}_0}  \right)=\frac{1}{Q^{\scriptscriptstyle{\rm DL}}_0}+\frac{G^{\scriptscriptstyle{\rm DL}}}{Q^{\scriptscriptstyle{\rm DL}}Q_0^{\scriptscriptstyle{\rm DL}}}\,.
$$
This leads to the following expansion
$$
\frac{1}{Q^{\scriptscriptstyle{\rm DL}}}=\frac{1}{Q^{\scriptscriptstyle{\rm DL}}_0}\sum_{k=0}^\infty \left(\frac{G^{\scriptscriptstyle{\rm DL}}}{Q^{\scriptscriptstyle{\rm DL}}_0}\right)^k\,.
$$
This expansion is uniformly convergent if $\alpha\geq\alpha_0>\alpha_0^{DL}$ and for all $\omega\in[-\pi,\pi]$ because $\left|G^{\scriptscriptstyle{\rm DL}}/Q^{\scriptscriptstyle{\rm DL}}_0\right|<c_0<1$  (where $c_0\thickapprox 0.87$). As a consequence of this we can write
\begin{equation}
Z^{\scriptscriptstyle{\rm DL}}(\alpha)=\frac{1}{2\pi}\int_{-\pi}^\pi\frac{\mathrm{d}\omega}{Q^{\scriptscriptstyle{\rm DL}}_0(\alpha,\omega)}
+\frac{1}{2\pi}\sum_{k=1}^\infty\int_{-\pi}^\pi\frac{(G^{\scriptscriptstyle{\rm DL}}(\alpha,\omega))^k}{(Q_0^{\scriptscriptstyle{\rm DL}}(\alpha,\omega))^{k+1}}\mathrm{d}\omega\,.
\label{Z}
\end{equation}
Only the first integral diverges when $\alpha\rightarrow\alpha_0^{\scriptscriptstyle{\rm DL}+}$ and, in fact, it can be computed in close form by using
$$
\frac{1}{2\pi}\int_{-\pi}^\pi \frac{\mathrm{d}\omega}{C^2(\alpha)+B^2(\alpha)\sin^2(\omega/2)}=\frac{1}{|C(\alpha)|\sqrt{C^2(\alpha)+B^2(\alpha)}}\,.
$$
When $\alpha\rightarrow\alpha_0^{\scriptscriptstyle{\rm DL}+}$ we have that
$$
C(\alpha)=\frac{1}{\sqrt{Z_*^{\scriptscriptstyle{\rm DL}}(\alpha)}}\sim\sqrt{\frac{\alpha-\alpha_0^{\scriptscriptstyle{\rm DL}}}{q_{*-1}^{\scriptscriptstyle{\rm DL}}}}
+\sum_{n=1}^\infty c_n(\alpha-\alpha_0^{\scriptscriptstyle{\rm DL}})^{n+\nicefrac{1}{2}}\,,
$$
and, hence,
$$
\frac{1}{2\pi}\int_{-\pi}^\pi \frac{\mathrm{d}\omega}{C^2(\alpha)+B^2(\alpha)\sin^2(\omega/2)}\sim\frac{\sqrt{q_{*-1}^{\scriptscriptstyle{\rm DL}}}}{|B(\alpha_0^{\scriptscriptstyle{\rm DL}})|}\frac{1}{(\alpha-\alpha_0^{\scriptscriptstyle{\rm DL}})^{\nicefrac{1}{2}}}+\sum_{n=0}^\infty d_n(\alpha-\alpha_0^{\scriptscriptstyle{\rm DL}})^{n+\nicefrac{1}{2}}\,.
$$
In the previous two expressions $c_n$ and $d_n$ are real coefficients. The remaining integrals in (\ref{Z}) can be seen to have the form
\begin{equation}
\frac{1}{2\pi}\int_{-\pi}^\pi \frac{1}{(Q^{\scriptscriptstyle{\rm DL}}_0(\alpha,\omega))^{k+1}}\left(\sin^{4k}\frac{\omega}{2}\right)\sum_n g_n(k,\alpha)P_n\big(\sin^2\frac{\omega}{2}\big)\,\mathrm{d}\omega
\label{form}
\end{equation}
for some regular functions $g_n$ and polynomials $P_n$. This can be seen by expanding
$$
\sin^2\frac{k\omega}{2}=\frac{1}{2}(1-\cos k\omega)=\frac{1}{2}(1-T_k(\cos\omega))=\frac{1}{2}\Big(1-2T_k\big(1-\sin^2\frac{\omega}{2}\big)\Big)\,
$$
in terms of the Tchebycheff polynomials $T_k$ and using the fact\footnote{This can be proved by using the following explicit expansion $$
T_k(x)= \sum_{k=0}^{\lfloor n/2\rfloor} \binom{n}{2k} (x^2-1)^k x^{n-2k}
$$} that the lowest degree monomial of $(1-T_k(1-2x^2))/2$ is $k^2x^2$. We need to use now
\begin{eqnarray}
&&\frac{1}{2\pi}\int_{-\pi}^{\pi}\frac{\sin^{2n}(\omega/2)}{(C^2+B^2\sin^2(\omega/2))^m}\mathrm{d}\omega\label{integralessin}\\
&&=\frac{1}{B^{2n}}\sum_{k=0}^n
\binom{n}{k}(-1)^{n-k}C^{2(n-m)}\left(1+\frac{B^2}{2C^2}\right)^{k-m}\!
_2F_1\left(\frac{1-k+m}{2},\frac{m-k}{2};1;\frac{B^4}{(2C^2+B^2)^2}\right)\,.\nonumber
\end{eqnarray}
Notice that, in our case we will always have $n\geq m$ owing to the presence of the term $\sin^{4k}(\omega/2)$ in (\ref{form}). The hypergeometric functions appearing in the previous expression are of the form
$$
F_N(z):=\,_2F_1\left(\frac{N+1}{2},\frac{N}{2};1;z\right)\,.
$$
It is straightforward to see that $F_N$ is a polynomial in the variable $z$ for $N\leq0$ and satisfies
$$
F_N(z)=\frac{F_{1-N}(z)}{(1-z)^{N-1/2}}\,, \quad {\rm for} \quad N=1,2,\ldots
$$
It is possible to show that (\ref{integralessin}) are always analytic functions in $C$ in a neighborhood of $C=0$ when $n\geq m$. Indeed, if the index $k$ in the sum satisfies $k\geq m$ the corresponding term in the r.h.s. of (\ref{integralessin}) is the product of $C^{(2n-k)}$ and an analytic function in $C$ because the hypergeometric function is a polynomial. For the terms with $m>k$ the hypergeometric function behaves as $1/C^{2(m-k)-1}$ and hence the full summand is of the form $C^{2(n-m)+1}$ times an analytic function in $C$. We then conclude that (\ref{integralessin}) are analytic functions of $\sqrt{\alpha-\alpha_0^{DL}}$ and then
$Z^{\scriptscriptstyle{\rm DL}}(\alpha)$ must have the form
$$
Z^{\scriptscriptstyle{\rm DL}}(\alpha)\sim \frac{q^{\scriptscriptstyle{\rm DL}}_{-\nicefrac{1}{2}}}{\sqrt{\alpha-\alpha^{\scriptscriptstyle{\rm DL}}_0}}+\sum_{n=0}^\infty q^{\scriptscriptstyle{\rm DL}}_{\nicefrac{n}{2}} (\alpha-\alpha^{\scriptscriptstyle{\rm DL}}_0)^{n/2}\,,\quad \alpha\rightarrow \alpha^{\scriptscriptstyle{\rm DL}+}_0\,,
$$
with
$$
q^{\scriptscriptstyle{\rm DL}}_{-\nicefrac{1}{2}}=\frac{\sqrt{q_{*-1}^{\scriptscriptstyle{\rm DL}}}}{|B(\alpha_0^{\scriptscriptstyle{\rm DL}})|}\,.
$$
In practice some of the next order terms in the expansion for $Z^{\scriptscriptstyle{\rm DL}}(\alpha)$ can be computed directly without using the series expansion introduced above. For example it is straightforward to see that
$$
q_0^{\scriptscriptstyle{\rm DL}}=\frac{1}{2\pi}\int_{-\pi}^{\pi}\left(\frac{1}{Q^{\scriptscriptstyle{\rm DL}}(\alpha_0^{\scriptscriptstyle{\rm DL}},\omega)}-
\frac{1}{Q^{\scriptscriptstyle{\rm DL}}_0(\alpha_0^{\scriptscriptstyle{\rm DL}},\omega)}\right)\mathrm{d}\omega\,.
$$
The asymptotic behavior of
$$
Z^{\scriptscriptstyle{\rm ENP}}(\alpha)=\frac{1}{\pi}\int_{-\pi}^{\pi}\frac{\sin^2 \omega\,\mathrm{d}\omega}{1-\sum_{k=1}^\infty e^{-\alpha \sqrt{k(k+2)}}\sin\big( (k+1)\omega\big)/\sin\omega}
$$
can be discussed along similar lines so we give only the most relevant steps. In this case we can write
$$
Z^{\scriptscriptstyle{\rm ENP}}(\alpha)=\frac{1}{\pi}\int_{-\pi}^{\pi} \frac{\sin^2\omega}{Q^{\scriptscriptstyle{\rm ENP}}(\alpha,\omega)}\mathrm{d}\omega
$$
with
$$
Q^{\scriptscriptstyle{\rm ENP}}(\alpha,\omega):=1-\sum_{k=1}^{\infty}\frac{\sin(k+1)\omega}{\sin\omega}e^{-\alpha\sqrt{k(k+2)}}\,.
$$
We introduce now
$$
Q^{\scriptscriptstyle{\rm ENP}}_0(\alpha,\omega):=\left(1-\sum_{k=1}^\infty (k+1)e^{-\alpha\sqrt{k(k+2)}}\right)+\left(\frac{2}{3}\sum_{k=1}^\infty k(k+1)(k+2)e^{-\alpha\sqrt{k(k+2)}}\right)\sin^2\frac{\omega}{2}
$$
and
$$
G^{\scriptscriptstyle{\rm ENP}}(\alpha,\omega):=Q^{\scriptscriptstyle{\rm ENP}}_0(\alpha,\omega)-Q^{\scriptscriptstyle{\rm ENP}}(\alpha,\omega)\,,
$$
which, as in the case of $Z^{\scriptscriptstyle{\rm DL}}$, is actually proportional to $\sin^4 (\omega/2)$. Now we can write the expansion
$$
Z^{\scriptscriptstyle{\rm ENP}}(\alpha)=\frac{1}{\pi}\sum_{k=0}^\infty\frac{(G^{\scriptscriptstyle{\rm ENP}}(\alpha,\omega))^k\sin^2\omega}{(Q^{\scriptscriptstyle{\rm ENP}}_0(\alpha,\omega))^{k+1}}\mathrm{d}\omega
$$
and show that it is an analytic function of $\sqrt{\alpha-\alpha_0^{\scriptscriptstyle{\rm ENP}}}$ by following the same procedure that we used for $Z^{\scriptscriptstyle{\rm DL}}$. Though the previous series provides a way to compute the coefficients of the power series for $Z^{\scriptscriptstyle{\rm ENP}}(\alpha)$ it is better, in practice, to compute them directly. In this case we have, for example,
$$
q_0^{\scriptscriptstyle{\rm ENP}}=\frac{1}{\pi}\int_{-\pi}^{\pi}\frac{\sin^3\omega}{\sum_{k=1}^\infty \big((k+1)\sin\omega-\sin(k+1)\omega\big)e^{-\alpha^{\scriptscriptstyle{\rm ENP}}_0\sqrt{k(k+2)}}}\mathrm{d}\omega\,.
$$


%

\end{document}